\documentclass[12pt]{article}
\usepackage{bbm} 
\usepackage{bm}

\usepackage{exscale}

\font\fourteenbf=cmbx12 scaled\magstep1

\def\a0size{6}

\renewcommand{\Im}{{\rm Im}}

\newcommand{\mphi}{{m}}

\newcommand{\lsi}{\raise0.3ex\hbox{$<$\kern-0.75em\raise-1.1ex\hbox{$\sim$}}}
\newcommand{\gsi}{\raise0.3ex\hbox{$>$\kern-0.75em\raise-1.1ex\hbox{$\sim$}}}
\newcommand{\lsim}{\mathop{\lsi}}
\newcommand{\gsim}{\mathop{\gsi}}
\renewcommand{\vec}[1]{{\bm #1}}

\begin{document}

\setlength{\baselineskip}{0.6cm}
\newcommand{\figysize}{16.0cm}
\newcommand{\figtopspace}{\vspace*{-1.5cm}}
\newcommand{\figbottomspace}{\vspace*{-5.0cm}}
  
% don't use the following line with revtex
%\renewcommand{\theequation}{\thesection.\arabic{equation}}
%\newcounter{saveeqn}
%\newcommand{\alphaeqn}{\refstepcounter{equation}\setcounter{saveeqn}{\value{equation}}%
%\setcounter{equation}{0}%
%\renewcommand{\theequation}{%
%        \mbox{\thesection.\arabic{saveeqn}\alph{equation}}}}%
%\newcommand{\reseteqn}{\setcounter{equation}{\value{saveeqn}}%
%\renewcommand{\theequation}{\thesection.\arabic{equation}}}

\begin{titlepage}
\begin{flushright}
BI-TP 2006/12\\
CERN-PH-TH/2006-077
\\
\end{flushright}
\begin{centering}
\vfill

{\fourteenbf \centerline{ 
Moduli decay in the hot early Universe
}
\vskip 2mm \centerline{ 
}}

\vspace{1cm}

Dietrich B\"odeker \footnote{e-mail: bodeker@physik.uni-bielefeld.de}

\vspace{.6cm} { \em 
Physics Dept., Theory Unit, CERN, CH--1211 Geneva 23, Switzerland
\\
and
\\
Fakult\"at f\"ur Physik, Universit\"at Bielefeld, D-33615 Bielefeld, Germany 
\footnote{Permanent address}
}

\vspace{2cm}
 
{\bf Abstract}

\end{centering}
 
\vspace{0.5cm}

We consider moduli fields interacting with thermalized relativistic
matter. We determine the  temperature dependence of their damping rate and find that
it is dominated by thermal effects in the high temperature regime, i.e. for
temperatures larger than their
mass. For a simple scalar model the damping rate is expressed through  the
known matter bulk viscosity. 
The high temperature damping rate is always smaller than the Hubble rate, so
that thermal effects are not sufficient for solving the  cosmological moduli
problem.  

\noindent

\vspace{0.5cm}\noindent

%PACS numbers: 
%Keywords: 
 
\vspace{0.3cm}\noindent
 
\vfill \vfill
\noindent
 
\end{titlepage}
 
%%%%%%%%%%%%%%%%%%%%%%%%%%%%%%%%%%%%%%%%%%%%%%%%%%%%%%%%%%%%%%%%%
%\section{Introduction}
%\label{sc:introduction} 
%%%%%%%%%%%%%%%%%%%%%%%%%%%%%%%%%%%%%%%%%%%%%%%%%%%%%%%%%%%%%%%%%

Many theories beyond the Standard Model of particle physics, such as string and
supersymmetric theories, 
contain directions in field space that have a flat
potential in the supersymmetric limit and couple to other fields only through
Planck scale suppressed interactions. 
If these directions are stabilized by
the same mechanism which is responsible for supersymmetry breaking, a particle
whose mass is of the 
order of a TeV and a dangerously large lifetime may result. Such
fields are often called moduli. If one of the moduli starts at a value
displaced from the minimum of the effective potential, 
it will perform coherent oscillations around its minimum. The energy of these
oscillations decreases
like the energy density in non-relativistic matter with the expansion of the
Universe, $ \rho \propto a ^{-3} $, 
while for relativistic matter  $ \rho  \propto a ^{-4 } $. Therefore the
moduli may soon contribute significantly to the total energy density of the
Universe. 

In the vacuum the decay rate of a  moduli field with mass $ \mphi  $ 
is  of order $   \mphi  ^ 3/ M  ^ 2$, where $ M = 2 \times 10 ^{18} $ GeV is
the reduced Planck mass \cite{coughlan}. The corresponding long lifetime $ \tau  =
\Gamma  ^{-1} \sim 
10 ^{5} $ $ ( {\rm TeV}/\mphi  ) ^ 3 $~s causes the cosmological moduli
problem \cite{coughlan}-\cite{randall}. 
%\cite{coughlan,goncharov,german,ellis,banks,randall}. 
Unless the initial amplitude of the moduli is very small, they
dominate the energy density of the Universe at some point, spoiling the
success of standard big bang nucleosynthesis.  

Many solutions to this problem have been suggested. Recently it was pointed
out in Ref.~\cite{yokoyama} that
in the hot early Universe the 
moduli oscillations could be damped much more strongly than in the vacuum. It
was found that the damping is in fact so strong that the moduli would
adiabatically follow the minimum of the effective potential,  making their
relic density harmless for cosmology \cite{yokoyama}. 
In this note we reconsider the thermal damping of scalar moduli
fields. Unfortunately, we cannot confirm the result of Ref.~\cite{yokoyama}. 

A modulus field   starts oscillating 
once the Hubble rate $ H $ drops below its mass $ \mphi  $,  which
happens at the temperature $ T _{\rm osc}\sim \sqrt{\mphi  M } \sim
 10 ^{10} $ $ {\rm GeV} \sqrt{m/{\rm TeV}}$.  The thermal 
corrections to the mass have a magnitude $ \lsim  T ^ 2 /M $,  which is of the same order
as $ H $ if the Universe is radiation-dominated. Thus, unless they come with a very
large coefficient \cite{lindeRelaxing}, they do not affect the parametric size
of $ T _{\rm osc} $,  and they can be neglected for $ T \ll
T _{\rm   osc} $. 

Thermal damping effects can become important if the
damping rate becomes larger than the Hubble rate. In order to see whether this
happens, 
we can neglect the expansion of the Universe. We consider temperatures $ T \gg
\mphi  $, because, as we shall see, only in this case does thermal
damping dominate over the vacuum damping. 
Then the matter fields, i.e.
fields that have unsuppressed interactions,  evolve on time scales of order $ T
^{-1} $, while the modulus field evolves much more slowly.

To describe the time evolution of this non-equilibrium system,
we imagine starting from  thermal equilibrium 
at $ t = - {\infty  }  $.  For simplicity the zero of the mo\-dulus
field $ \varphi  $ is
chosen such that  
its expectation value at the considered temperature vanishes. To drive the
system out of equilibrium we add the  term  $ { \cal
  L } _{\rm ext} = j \varphi $  to the Lagrangian,  where  $
j $ is an external c-number field that vanishes for $ t \to -{\infty  }
$. Then, at some finite time, $ j $ is switched off again. From then on we have
an isolated system which is not in thermal equilibrium. 
The way the system got out of
equilibrium does not matter because 
$ \varphi  $  evolves much more slowly than
all other degrees of freedom, since $ m \ll T $. On time scales over which $
\varphi  $ 
changes, the other fields have long come into thermal
equilibrium and do not remember anything about their past. 

As long as $ \langle \varphi  \rangle \ll M$ one can use the linear
approximation to compute the expectation value of $ 
\varphi  $ \cite{landau5}: 
\begin{eqnarray} 
  \langle \varphi  ( x ) \rangle \simeq  \int d ^ 4 x \Delta  _{\rm ret} ( x - x' )
  j ( x' )
\end{eqnarray} 
with the retarded propagator 
\begin{eqnarray} 
  \Delta  _{\rm ret} ( x - x' )
  \equiv \langle \varphi  ( x ) \varphi  ( x' ) \rangle _{\rm ret} 
  \equiv i \Theta  ( t - t' ) \left \langle [ \varphi  ( x ) , \varphi  ( x' ) ] 
  \right \rangle _{\rm eq}
   . 
\end{eqnarray} 
Here the expectation value is taken in a thermal ensemble with the initial
temperature~\footnote{The system may get heated by applying the external
  field, but this effect does not contribute in the linear approximation.}.
The Fourier transform of the retarded propagator has an analytic continuation
\begin{eqnarray}
  \Delta ( p )  = \frac{ 1 } { - p ^ 2 + \mphi  ^ 2 + \Pi  ( p ) }
\end{eqnarray} 
into the upper half of the complex frequency plane such 
that
\begin{eqnarray} 
  \Delta  _{\rm ret} ( p ) = \Delta  ( p ^ 0 + i \epsilon  , \vec p )
  .
\end{eqnarray} 
We are interested in spatially homogeneous oscillations and can therefore
consider an $ \vec x $-independent $ j $. 
Without any interaction $ \varphi  $ would simply oscillate with the frequency
$\mphi  $. With interaction, $ \mphi ^ 2 $ receives 
finite temperature corrections  given by the real part of $ \Pi  $.
The  imaginary part of $ \Pi  $ leads to exponential damping of the $
\varphi  $ oscillations. We assume that the damping rate is much smaller than
the oscillation frequency,  which
will be verified a posteriori \footnote{Yokoyama \cite{yokoyama} finds a
  damping rate which is much larger than the frequency.}.
Then  the  damping rate  is given by 
\begin{eqnarray} 
  \gamma  = - \frac{ 1 }  { 2 m _ {  T }   }
   \,\Im \, \Pi  ( m _ {  T }  + i \epsilon   , \vec 0) 
   , 
\end{eqnarray} 
where $ m _ {  T } $ is the thermally corrected frequency.
Since $ m _ {  T } $ is much smaller than all other relevant energy scales in
the problem, we can write 
\begin{eqnarray} 
  \gamma \simeq 
  - \, \lim _ { \omega   \to 0} \,\frac{ 1 } { 2 \omega  } 
   \, \Im \, \Pi  (\omega  + i \epsilon   , \vec 0  ) 
   .
   \label{p10} 
\end{eqnarray}
This limit exists since $ \Pi  $ is real at zero frequency.  
Therefore, at leading order in $ \mphi  /T  $, the damping rate 
does not depend on $ \mphi  $ 
\footnote {In Ref.~\cite{yokoyama} an $ \mphi  $-dependent
damping rate was obtained.
One cause for this discrepancy appears to be that in going from Eq.~(21) to 
 Eq.~(22) of
  Ref.~\cite{yokoyama} the $ \chi  $-damping rate $ \Gamma  _ p $  has been
  neglected in 
  favour of $ \mphi  $,  even though $ \Gamma  _ p \gg \mphi  $.
 }. 
For large $ T $ one can neglect the masses of the matter particles so that, 
by dimensional analysis,
\begin{eqnarray}
  \gamma  = f  \frac{ T ^ 3 } { M ^ 2 }
  ,
  \label{gamma} 
\end{eqnarray} 
where  $ f $ is a function of the (dimensionless) gauge, Yukawa, and scalar couplings. 
For $ T \lsim   T _{\rm osc} $ we have $ \gamma  \lsim  m f \sqrt{m/M}  $, which
is indeed much smaller than $ m $. 

In general the calculation of $ f $ is non-trivial. We consider a simple model
\cite{yokoyama}  where 
$ \varphi  $ is coupled to a massless scalar field $ \chi   $, 
which mimics a Standard Model field. We
will see that for this case the damping rate 
can be expressed through the bulk viscosity, which has been computed
elsewhere. The Lagrangian is 
\begin{eqnarray} 
    { \cal L } = { \cal L } _ \varphi + { \cal L } _ \chi  + { \cal L }
  _{\varphi  \chi   }  
  , 
  \label{model} 
\end{eqnarray} 
with 
\begin{eqnarray} 
  { \cal L } _ \varphi = \frac12  \left ( \partial \varphi  \right ) ^ 2 
  - \frac{ \mphi  ^ 2}{2}
  \varphi  ^ 2 
\end{eqnarray}
\begin{eqnarray} 
  { \cal L } _ \chi   =   \frac12   \left  ( \partial \chi  \right ) ^ 2  
  - \frac{ \lambda   }{4!} \chi  ^ 4
\end{eqnarray}
\begin{eqnarray} 
  { \cal L }   _{\varphi  \chi  } = 
  \frac{ g ^ 2 }{2M } \varphi   \left  ( \partial \chi  \right ) ^ 2  
  .
\end{eqnarray} 
The coupling constant $ \lambda   $  is supposed to be reasonably large, so that 
the field $ \chi  $ is in thermal equilibrium, but small enough so that
perturbation theory can be applied.  At leading order in $ M ^{-1}  $ the $
\varphi  $ self-energy can be written as 
\begin{eqnarray}
    \Pi  ( p ^ 0 + i \epsilon , \vec p  ) 
    = 
    - \frac{ g ^ 4 } { 4 M ^ 2 } \int d ^ 4 x e ^{i p \cdot x } 
   \left \langle 
     \left (  \partial \chi  \right )  ^ 2 ( x )
     \left (  \partial \chi   \right )  ^ 2 ( 0 )
   \right \rangle 
_{\rm ret} 
  , 
  \label{p23} 
\end{eqnarray} 
where the 2-point function in  Eq.~(\ref{p23}) 
has to be evaluated for $ M \to {\infty  } $. In this limit the operator $ ( 
\partial \chi  ) ^ 2 $ is related to the trace of the energy momentum tensor
of $ \chi  $ 
\begin{eqnarray}
  T  ^ { \mu  \nu  } 
  \equiv 
  \partial ^ \mu  \chi  \partial ^ \nu  \chi  - \eta
  ^{\mu  \nu  } { \cal L _ \chi   }
\end{eqnarray} 
through the trace anomaly \cite{collins}, 
\begin{eqnarray} 
  T ^ \mu  { } _ \mu  =  - \frac{ \beta  ( \lambda  ) } { 4 \lambda  } \left (
    \partial \chi  \right ) ^ 2
  + \cdots   
  .
  \label{p23.3a} 
\end{eqnarray} 
Here $ \beta  ( \lambda  ) = 3 \lambda  ^ 2/( 16 \pi  ^ 2 ) + O ( \lambda  ^ 3
) $ is the $ \beta  $ function of the scalar $ \chi  ^ 4 $ theory. The
ellipsis in Eq.~(\ref{p23.3a}) denotes total derivatives and terms which vanish
by the equations of motion, neither of which contribute to  the imaginary part
of $ \Pi  $ in the limit~(\ref{p10}). 
Combining Eqs.~(\ref{p10}),  (\ref{p23}), and (\ref{p23.3a}), we obtain at
leading order in $ \lambda  $ 
\begin{eqnarray}
  \gamma  
  = %
  \frac{ g ^ 4 }{8 M ^ 2 } 
  \frac{ ( 8 \pi  ) ^ 4 } { 9 \lambda  ^ 2 } 
  \lim _{\omega  \to 0} \frac{ 1 } {
  \omega  } {\rm Im} \int d ^ 4 x e ^{i \omega  x^0} \langle T ^ \mu  { } _ \mu 
  ( x ) T ^ \nu  { } _ \nu  ( 0 ) \rangle _{\rm ret} 
  .
  \label{p23.3b} 
\end{eqnarray} 
The right-hand side is related to the bulk viscosity $ \zeta  $ of a gas of $ \chi
$ particles through the Kubo relation \cite{jeon}  
\begin{eqnarray} 
  \zeta  = 
  \frac19
 \lim _{\omega  \to 0} \frac{ 1 } {
  \omega  } {\rm Im} \int d ^ 4 x e ^{i \omega x^0} \langle T ^ \mu  { } _ \mu 
  ( x ) T ^ \nu  { } _ \nu  ( 0 ) \rangle _{\rm ret} 
  .  
  \label{p25} 
\end{eqnarray} 
We thus find that the damping rate of the moduli field $ \varphi  $ in this
model is directly proportional to the bulk viscosity of the plasma 
\begin{eqnarray} 
  \gamma  
  = 
  \frac{  g ^ 4 } { 8 M ^ 2 } \, \frac{ ( 8 \pi  ) ^ 4 } { \lambda
  ^ 2 } \, \zeta 
  .
  \label{p25.2} 
\end{eqnarray} 
In order to compute $ \zeta  $ even at leading order in $ \lambda   $ 
one has to sum an infinite set of diagrams \cite{jeon}. This turns out to be 
equivalent to solving an appropriate Boltzmann equation, which accounts for 
particle-number-changing processes. The result is 
\cite{jeonYaffe} 
\begin{eqnarray} 
  \zeta  
  = % * 
  \frac{ b } {6 ( 32 \pi  ) ^ 4 } \lambda   \ln ^ 2( \xi  \lambda  ) T ^ 3
  , 
\end{eqnarray} 
where $ b = 5.5 \times 10 ^ 4 $ and $ \xi    =  \exp  \left ( 
    { 15 \zeta  ( 3 )} /{  \pi  ^ 2 } \right )/96 = 0.064736 $. Then 
we finally obtain 
\begin{eqnarray}
  \gamma  
%  = % *
%  \frac{ b } { 3 \times 2 ^{12}  } 
%  g ^ 4  \frac{\ln ^ 2( \xi  \lambda  ) }{\lambda
%  }  \frac{ T ^ 3 } { M ^ 2 }
  \simeq % *
  4.5 \,
   g ^ 4  \frac{ \ln ^ 2( \xi  \lambda  ) }{\lambda
  }  \frac{ T ^ 3 } { M ^ 2 }
  \label{p25.3.1} 
\end{eqnarray} 
for the $ \varphi  $ damping rate in the model (\ref{model}). 

We have thus seen that the damping rate of moduli fields in a thermal
environment   is  parametrically larger than in the vacuum,  when $ T
\gsim  \mphi  $. However, it is always small compared to the 
Hubble rate.  Therefore thermal effects alone cannot solve the cosmological
moduli problem. 

\vspace{.5cm}
{\bf Acknowledgements} 
I would like to thank  M.~Ratz   for bringing Ref.~\cite{yokoyama}  to my
attention, W.~Buchm\"uller,  S.~Huber, and  M.~Laine for useful
comments, and the referee for pointing out a numerical  error in an earlier
version of this paper. 
This work was supported in part  through the
DFG funded Graduate School GRK 881.

\end{document}